\documentclass[pra,twocolumn,showpacs,superscriptaddress,preprintnumbers,amsmath,amssymb,floatfix]{revtex4}
\usepackage{bm}
\usepackage{graphicx}
\usepackage{amsbsy}
\usepackage{amsmath}
\usepackage{amsfonts}
\usepackage{amsthm}
\usepackage{color}
\usepackage{mathrsfs}
\usepackage{graphicx}
\usepackage{dcolumn}
\usepackage{bm}

\newcommand{\bwt}{\begin{widetext}}
\newcommand{\ewt}{\end{widetext}}
\newcommand{\bea}{\begin{eqnarray}}
\newcommand{\eea}{\end{eqnarray}}

\begin{document}

\title{Atomic  quantum transistor based on swapping operation}

\author{Sergey A. Moiseev}
\email{samoi@yandex.ru}

\affiliation{Kazan Physical-Technical Institute of the Russian Academy of Sciences,
10/7 Sibirsky Trakt, Kazan, 420029, Russia}
\affiliation{Institute for Informatics of Tatarstan Academy of Sciences, 20 Mushtary, Kazan, 420012, Russia}


\author{Sergey N. Andrianov}
\affiliation{Institute for Informatics of Tatarstan Academy of Sciences, 20 Mushtary, Kazan, 420012, Russia}

\author{Eugene S. Moiseev}

\affiliation{Kazan Physical-Technical Institute of the Russian Academy of Sciences,
10/7 Sibirsky Trakt, Kazan, 420029, Russia}
\date{\today}


\begin{abstract}
We propose an atomic quantum transistor based on exchange by virtual photons between two atomic systems through the control \emph{gate-}atom. The quantum transistor is realized in two QED cavities coupled in nano-optical scheme. We have found novel effect in quantum dynamics of coupled three-node atomic system which  provides control-$SWAP(\theta)-$ processes in quantum transistor operation.
New possibilities of quantum entanglement in an example of bright and dark qubit states have been demonstrated for  quantum transport in the atomic chain.
Potentialities of the proposed nano-optical design for quantum computing and fundamental issues of multi-atomic physics are also discussed.

\end{abstract}



\maketitle
\emph{Inroduction:}
One of the basic elements of microelectronic logic devices is the transistor - a three-port device where the electron flux through two ports is controlled by the third port. The same is with optoelectronics where light is controlled by light via electrons or atomic medium \cite{Gibbs1985,Wakita1998} in integrated design \cite{Kwiat2008}. Moor's law dictates increasing of elements number per chip.  But confining of light on the subwavelength scale is complicated.  Therefore excitonic transistor where exciton fluxes of medium excitation are controlled directly by a gate voltage \cite{High2007,High2008,Grosso2009} without using light itself for logic operations seems to be a promising approach.
As for quantum logics \cite{Nielsen2000} rapidly developing last decades, it also strongly demands for creation of transistors like in the case of its classical counterpart. Several quantum transistor designs were proposed \cite{Kane1998,Vrijen2000,Duan2004,Chang2007,Shen2007,Birnbaum2008,Hwang2009,Mucke2010}. Among them, spintronics approach \cite{Kane1998,Vrijen2000} is interesting from dense integration point of view. It is useful for creation of NMR quantum computer in microwave range.
Here, we propose quantum transistor in atomtronics approach based on  $SWAP-$ operation using exchange between atoms by virtual photons that was proposed initially in \cite{Imamoglu1999,Schuch2003,Majer2007} and observed experimentally for neutral atoms in traps recently \cite{Anderlini2007,Lin2009,Lundblad2009}. This transistor does not use real photons and therefore permits subwavelength integration extremely urgent for quantum technologies dealing with single atoms or mesoscopic atomic ensembles. We show that such kind of transistors can be used for realization of control $SWAP-$ and $\sqrt{SWAP}-$ processes. Finally we discuss possible applications of the atomic transistor in optical solid state quantum computer and fundamental issues of quantum nature revealed in multi-particle systems.

\emph{Basic scheme:}
The atomic quantum transistor (AQT)
contains
pair of resonant \emph{multi-atomic ensembles} (or pair of two-level atoms, we call it ME1 and ME2-nodes)
and a single three-level \emph{gate-}atom.
ME1 and ME2-nodes are characterized by long-lived coherence  and coupled with each other via the \emph{gate}-atom which  could be a natural or artificial atom with appropriate energy levels and transition dipole moments. Depending on the initial state, \emph{gate}-atom controls (switches on in the state $|0 \rangle_0$ or blocks in the state $|b\rangle_0$ ) an exchange by virtual photons between the two atomic nodes providing AQT operation in two interesting regimes. In the first regime of AQT, only a control-$SWAP$ operation can be implemented, the second regime of AQT provides besides a control-$\sqrt{SWAP}$ operation for quantum states of the atomic nodes.

\begin{figure}
\includegraphics[width=0.5\textwidth,height=0.2\textwidth]{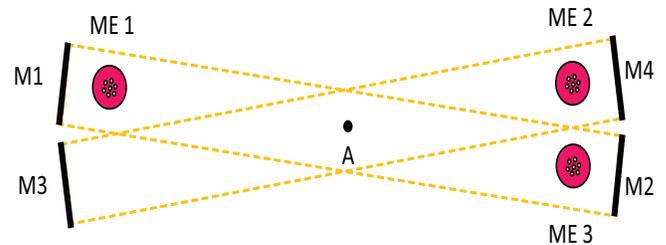}
\caption{Nano-optical scheme of atomic quantum transistor. Two QED-cavities contain three atomic ensembles- ME1,ME2, ME3 and single  \emph{gate}-atom "A" ; two field modes of the cavities are coupled with other through the \emph{gate-}atom; M1-M4 are mirrors. }
\label{Figure1}
\end{figure}

The proposed AQT exploits nano-optical scheme of the light-atom interface depicted in Fig.\ref{Figure1}. ME1 and ME2-nodes are situated in different single mode resonators.
The resonators can be fabricated with nano-optical waveguide technology.
Such type of the nano-optical fiber  with enhanced evanescent field has been demonstrated recently \cite{Vetsch2010} as a promising method for strong coupling of evanescent light field with single atoms situated at small distance ($\approx 200$ nm) from the fiber. There, the authors also observed  that the trapped atoms do not acquire essential line broadening.
Thus ME1 and ME2 nodes could be the atomic systems trapped close to the surface of the waveguide resonators or it could be rare-earth ions  in the inorganic crystal (and NV centers in nano-diamond) embedded in a body of the waveguide resonators.
We put the \emph{gate}-atom in the crossing point of the resonator axis as depicted in Fig. \ref{Figure1}. Here, \emph{gate}- atom will interact intensively with each field mode. At the same time the field modes will be  decoupled from each other for  non-collinear axis orientation of the QED-cavities.

Let us take that working frequencies of the atoms 
shifted from the frequency $\omega_e$ of cavity field modes where
we also assume equalized frequencies of the field modes $ \omega _{k_\alpha }=\omega _{e}$ ($\alpha=1,2$ is an index of the modes,
 \textit{$k_{\alpha} $} are wave vectors of the field modes).  The effective Hamiltonian describing the process can be obtained by unitary
transformation of initial James-Cummings model Hamiltonian that excludes
non-resonant atomic-photon interaction in the first perturbation order
\cite{Schuch2003,Moiseev2011}. This procedure yields the following Hamiltonian of the ME1 and ME2-nodes and \emph{gate}-atom in a photon vacuum state:

\begin{align}
\label{main Hamiltonian1}
 H_s =H_{(s,o)}+V,
\end{align}

\noindent

\begin{align}
\label{main Hamiltonian_so}
 H_{(s,o)} =&\hbar
\tilde\omega _{0} S_{c }^z+\hbar \omega_1 \sum\limits_{j_1 }^{N_1 } {S_{j_1 }^z } +\hbar \omega_2
\sum\limits_{j_2 }^{N_2 } {S_{j_2 }^z }
\nonumber       \\
 &+\hbar \sum\limits_{\alpha=1,2 }
\sum\limits_{j_{\alpha}, j_{\alpha}' }^{N_{\alpha}, N_{\alpha} }
 { \frac{ {g_{j_{\alpha }}g_{j_{\alpha }'} ^{\ast}} }{ \Delta _{\alpha} }
 e^{i\vec {k}_{\alpha} \vec {r}_{(j_{\alpha}, j_{\alpha}' )}}S_{j_{\alpha} }^+ S_{j_{\alpha}' }^- },
\end{align}

\begin{align}
\label{main Pertubation}
V=\hbar\sum\limits_{\alpha=1,2 }\sum\limits_{j_{\alpha}}^{N_{\alpha} }
{\left\{
\frac{ {g_{j_{\alpha }}g_{\alpha } ^{\ast}} }{ 2 }
(\frac{ {1} }{ \Delta _{\alpha} }+\frac{ {1} }{ \Delta _{0} })
{e^{i\vec {k}_{\alpha } \vec {r}_{(j_{\alpha }, a) }
}S_{j_{\alpha } }^+ S_{c }^- +h.c. } \right\}},
\end{align}

\noindent
where  $H_{(s,o)}$ describes Hamiltonian of ME1, ME2-nodes and \emph{gate-}atom, while $V$ determines an exchange processes between these three atomic groups, $\tilde\omega _0= \omega _0+
\frac{{1}}{\Delta_0} \left(\left|{\tilde g_{1 } } \right|^2+\left|{\tilde g_{2 } } \right|^2 \right)$;
$\omega _0$ and  $\omega_1$, $\omega_2$ are the initial transition frequencies of \emph{gate-}atom and the atoms in two nodes; $N_{1}$ and $N_{2}$ are
the quantity of atoms in ME1 and ME2-nodes; $S_{j_1 }^z $, $S_{j_2 }^z $ and
$S_{c }^z $ are z-components of effective $1/2-$spin
 in sites $j_{1}$, $j_{2}$ of
ME1, ME2 and for gate-atom; $S_{j_1 }^\pm $, $S_{j_2 }^\pm $ and $S_{a
}^\pm $ are an appropriate raising and lowering $1/2-$spin operators;
$g_{j_{\alpha }}$ are the photon-atom coupling for $j_{\alpha }$ atom in ${\alpha }$ atomic node,
(\textit{$\alpha $} = 1,2)
$\tilde g_{\alpha }$ are the coupling constants of gate-atom with photon in  $\alpha$-evanescent mode
 ; $\Delta _\alpha =\omega _\alpha -\omega _{e} $,
 $\Delta _o =\omega _o -\omega _{e } $ are frequency offsets;
  $r_{(ij)} $ is radius-vector connecting sites $i$ and $j$, where $r_{a}=0 $ is used.

\emph{Swap-dynamics:}
Let's consider double $SWAP$-processes. We assume that ME1-node is prepared in the single atomic excitation state, while the second ME2 node and \emph{gate-}atom stay in  the ground states that gives the following initial quantum state
$|\Psi _1 \rangle=|0\rangle_0 |1\rangle_1  |0\rangle_2\equiv |0\rangle_0 |10\rangle$ (where
$|a\rangle_1 |b\rangle_2 ... |f\rangle_n\equiv|ab...f\rangle$),
$\left| 1 \right\rangle _\alpha =\frac{1}{\sqrt {A_\alpha }
}\sum\limits_{j_\alpha }^{N_\alpha } {g_{j_\alpha } e^{ik_\alpha r_{j_\alpha
} }\left| {0_1 } \right\rangle \left| {0_2 } \right\rangle ...\left| 1_{j_\alpha }
\right\rangle ...\left| {0_{N_\alpha } } \right\rangle } $ is a
the  single atomic excitation state where
$A_\alpha =\sum\limits_{j_\alpha }^{N_\alpha } {\vert g_{j_\alpha } \vert ^2}$
and
$\left| 0 \right\rangle _\alpha =\left| {0_1 } \right\rangle \left| {0_2 }
\right\rangle ...\left| {0_{N_\alpha } } \right\rangle $ is a ground
state in $\alpha $- atomic node.
 The studied quantum dynamics of the atoms described by the effective Hamiltonian (\ref{main Hamiltonian1}) yields the wave function
\begin{align}
\label{wave function}
|\Psi ( t )\rangle=c_0(t)|\Psi _0 \rangle+c_1(t)|\Psi _1 \rangle +c_2( t)|\Psi _2\rangle,
\end{align}
decomposed into the quantum superposition of single atomic excitation states in the system of \emph{gate-}atom plus ME1 and ME2-nodes,
where  $|\Psi _0 \rangle=  |1\rangle_0|00\rangle$,
$|\Psi _2 \rangle=|0\rangle_0 |01\rangle$,
 $\left| 0 \right\rangle _0 $ and $\left| 1
\right\rangle _0 $ are the ground and excited states of the \emph{gate-}atom.

Let us assume equal numbers  $N_1=N_2=N$ and resonant frequencies $\omega_1=\omega_2=\omega$ of atoms in ME1 and ME2 nodes. By taking into account
$H_{(so)} \left| {\Psi _0 } \right\rangle =(E_o +\hbar \tilde {\omega }_0)\left| {\Psi _0 } \right\rangle,$
$H_{(so)} \left| {\Psi _\alpha } \right\rangle =(E_o +\hbar \tilde\omega) \left| {\Psi _\alpha } \right\rangle$
and
$\left\langle {\Psi _0 } \right|V \left| {\Psi _\alpha } \right\rangle=\sqrt {N } \hbar\Omega _{\alpha c},$
(where
$E_o =-\frac{\hbar \tilde {\omega }_0 }{2}-N\hbar \omega,$ $\tilde\omega = \omega +N\Omega_\alpha, $
$\Omega _{\alpha c} =\frac{g_a \sqrt {\left\langle {\vert {g}_{j_\alpha} \vert ^2} \right\rangle } }{2}(\frac{1}{\Delta _\alpha }+\frac{1}{\Delta
_0 }),$
$\Omega _\alpha =\textstyle{1 \over {N_\alpha }}\sum\limits_{j_\alpha
}^{N_\alpha } {\frac{\vert g_{j_\alpha } \vert ^2}{\Delta _\alpha }}$
 ) we find a solution for the amplitudes $c_{0,1,2}(t)=e^{-iE_o t/\hbar}\tilde c_{0,1,2}(t)$ using the initial quantum state $c_0(t=0)=c_2(t=0)=1$ and $c_1(t=0)=1$:

\begin{align}
\label{main solution}
\tilde c_0 (t)=2i\frac{\sqrt {N_\alpha } \Omega _{\alpha c}}{S}e^{-i(\tilde {\omega } +\frac{\Delta }{2})t}\sin
(\frac{St}{2}),
\nonumber       \\
\tilde c_1 (t)=\frac{1}{2}e^{-i\tilde {\omega } t}\{1+e^{-i\frac{\Delta
t}{2}}[\cos (\frac{St}{2})-i\frac{\Delta }{S}\sin (\frac{St}{2})]\},
\nonumber       \\
\tilde c_2 (t)=\frac{1}{2}e^{-i\tilde {\omega } t}\{-1+e^{-i\frac{\Delta
t}{2}}[\cos (\frac{St}{2})-i\frac{\Delta }{S}\sin (\frac{St}{2})]\},
\end{align}
\noindent
where $S=\sqrt {8 N {\Omega_{\alpha c}}^2+\Delta ^2},$
$\Delta=\tilde {\omega }_0 -\tilde {\omega }.$

\noindent
The solution (\ref{main solution}) reveals an interesting dynamics in two particular cases.

\emph{Resonant swapping $(\Delta=\tilde {\omega }_0 -\tilde {\omega }=0)$:}
We assume that it is possible to control the atomic resonant frequencies $\tilde {\omega }_0$ and $\tilde {\omega }$
by using Stark or Zeeman effects in external electric or magnetic fields applied to the \emph{gate-}atom and ME1, ME2 -nodes. By equalizing the frequencies $\tilde {\omega }_0 =\tilde {\omega }$ in Eq. (\ref{main solution}), we find a nutation with a maximum oscillation amplitude in a system of three atomic systems:
\begin{align}
\label{main solution-a}
\tilde c_0 (t)=\frac{i}{\sqrt {2}}e^{-i\tilde {\omega } t}\sin(\frac{S_ot}{2}),
\nonumber       \\
\tilde c_1 (t)=e^{-i\tilde {\omega } t} \cos^2 (\frac{S_ot}{4}),
\nonumber       \\
\tilde c_2 (t)=-e^{-i\tilde {\omega } t} \sin^2 (\frac{S_ot}{4}),
\end{align}
\noindent
where $S_o=2 \sqrt {2 N }{\Omega_{\alpha c}}$.

\noindent
Eq. (\ref{main solution-a}) demonstrates a 100\% transfer of excitation from ME1 to ME2-node at the moment of time $t= t^{(r)}_{swap}=2\pi/S_o$ where $c_0 =c_1 =0$ and $|c_2 |=1.$ We note that  $|c_0 ( t)| $ gets a maximum $|c_0| =1/\sqrt{2}$ at $t"=t^{(r)}_{swap}/2$ that corresponds to 50\% of probability to find the excitation on \emph{gate-}atom. At this moment of time  $|c_1 ( t")| =|c_2 ( t")|=1/2$, i.e. other 50 \% of the excitation are spread equally  between the ME1 and ME2-nodes.

For arbitrary initial state of ME1-node
$|\psi(\phi_1)\rangle_1=\alpha_1 \left| 0 \right\rangle _1 +\beta_1 e^{i\phi_1}\left| 1 \right\rangle _1,$
we get

\begin{align}
\label{general_swap}
\left|0 \right\rangle _0
|\psi_1(\phi_1)\rangle_1
\left| 0 \right\rangle _2
\xrightarrow{SWAP}
\left| 0 \right\rangle _0
\left| 0 \right\rangle _1
|\psi(\phi_1-\tilde {\omega }t^{(r)}_{swap}+\pi)\rangle_2,
\end{align}

\noindent
with phase transformation $\phi_1\rightarrow\phi_1-\tilde {\omega }t^{(r)}_{swap}+\pi$.

\emph{Nonresonant swapping:}
More general scenarios of the excitation transfer occurs for nonresonant interaction with spectral detuning  $\Delta=2\sqrt {2 N/3 }{\Omega_{\alpha c}}= S/2$ between the resonant frequencies of \emph{gate-}atom and the atomic nodes.
By putting
$t^\ast=2 \pi/S=\frac{\pi}{2\Omega_ {\alpha c}} \sqrt{\frac{3}{2N}}\equiv    t_{\sqrt{swap}} $ in Eq.(\ref{main solution}),
we get  at this  moment of time

\begin{align}
\label{main solution-c}
|\Psi ( t^\ast )\rangle=  e^{-i\delta\phi}
\{|\Psi _1 \rangle+i |\Psi _2 \rangle\}
/{\sqrt{2}},
\end{align}

\noindent
where
$\delta\phi=(E_o/\hbar+\tilde {\omega }+\Delta/4) t_{\sqrt{swap}}-\pi/2$,
$ t_{\sqrt{swap}} $
corresponds to a $\sqrt {SWAP}$-operation (equal redistribution of the initial excitation in  ME1-node over the ME1 and ME2-nodes).

\begin{figure}
\includegraphics[width=0.4\textwidth,height=0.2\textwidth]{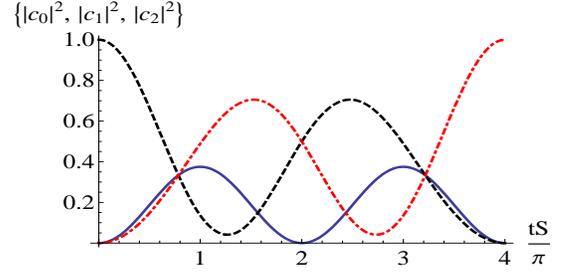}
\caption{Dynamics of the excitation probabilities in nonresonant swapping for \emph{gate-}atom ($|c_0(t)|^2$ blue solid  line), ME1-node ($|c_{1}(t)|^2$ black dashed line) and ME2-node ($|c_{2}(t)|^2$ red dot-dashed line).}
\label{Figure2}
\end{figure}

Succeeding atomic evolution yields a complete transfer of the state between ME1 and ME2-nodes  for the moment of time only $\sqrt{3}$ times longer than resonant swapping time $t^{(nr)}_{swap}=2t_{\sqrt{swap}}=\sqrt{3}t^{(r)}_{swap}$. Temporal behavior of excitation probabilities for \emph{gate-}atom and ME1, ME2-nodes is depicted in Fig. \ref{Figure2}. At the point $tS/\pi=2$, excitation probability density at central atom is zero when the excitation probabilities of ME1 and ME2-nodes are equal each other and equal $1/2$ that is we are in purely entangled state of spatially separated ensembles ME1 and ME2. It had become possible due to the quantum interference of the excitations in distant nodes at the special choice of parameters $\Delta=S/2$  when intranode detuning oscillation frequency and internode swapping frequency coincide. On the other hand there are also two points in Fig.\ref{Figure2} ($tS/\pi\approx 0.8$ and $tS/\pi\approx 3.2$) where excitation probabilities are equal each other and equal $1/3$ that is we are in purely entangled state of three spatially separated atomic systems - \emph{gate-}atom, ME1 and ME2-nodes.

By using the initial state  $|\psi(\phi_1)\rangle_1$, we get
\begin{align}
\label{main sqrt_SWAP}
&\left|0 \right\rangle _0
|\psi(\phi_1)\rangle_1
\left| 0 \right\rangle _2
\xrightarrow{\sqrt{SWAP}}
|0\rangle _0|\Phi^{(nr)}_{\sqrt{swap}}(\psi)\rangle,
\end{align}
\noindent

\begin{align}
\label{function of sqrt_SWAP}
|\Phi^{(nr)}_{\sqrt{swap}}(\psi)\rangle =
\alpha_1 \left| 00 \right\rangle
+ \beta_1 e^{i(\phi_1-\delta\phi)}
[|10 \rangle+i |01 \rangle]/{\sqrt{2}}.
\end{align}

 Nonresonant $SWAP$-gate is realized  with the same
phase transformation $\phi_1\rightarrow\phi_1-\tilde {\omega }t^{(nr)}_{swap}+\pi$ as
in resonant swapping. Moreover $\sqrt{SWAP}$ gate is directly generalized to $SWAP(\theta_{n,m})$-gate if we use a swapping time $t=2m \pi/S$  and swapping phase $\theta_{n,m}=\pi m/(4n)$ with spectral detuning $\Delta=S/(2n)$ where $m,n$ are integers. Here, $SWAP$-process is realized for $m=2n$ and $\sqrt{SWAP}$ occurs for $n=m$.

\emph{Control-$SWAP(\theta_{n,m})$-operations:}
Here, we take into account that \emph{gate-}atoms is a three level system and excitation in it can be transferred from the states $|0\rangle_0,|1\rangle_0$ to state $|b\rangle_0$  highly  decoupled from  ME1 and ME2-nodes. Availability  of the states $|0\rangle_0$ and $|b\rangle_0$ in the \emph{gate-}atom opens a possibility for realization of control-$SWAP(\theta_{n,m})-$  gates. Below we demonstrate this AQT working in a quantum fashion for control-$SWAP-$ and control-$\sqrt{SWAP}$ processes.

Initially  we transfer the quantum state of control ME3-node $|\Psi_c(\phi_c)\rangle_3=\alpha_c|0\rangle_3+\beta_c e^{i\phi_c}|1\rangle_3$ to the \emph{gate-}atom. For this step we equalize resonant frequency $\tilde\omega_{0}$ of the \emph{gate-}atom with the frequency of control two-level atoms $\tilde\omega_3$ providing a transfer of
$|\Psi_c(\phi_c)\rangle_3$
to \emph{gate-}atom

\begin{align}
\label{main controlo}
|0\rangle_0
|\Psi_c(\phi_c)\rangle_3
\xrightarrow{(g \leftrightarrow ME3)}
|\Psi_c(\phi_c+\pi/2)\rangle_0
|0\rangle_3.
\end{align}
\noindent
In the second step we apply an auxiliary laser $\pi$-pulse to \emph{ gate-}atom which transforms the state component $i\beta_c e^{i\phi_c}|1\rangle_0$  in Eq.(\ref{main controlo}) to the blockade state $-\beta_c e^{i\phi_c}|b\rangle_0$. By taking into account the initial state of ME1 and ME2-nodes, we get
\begin{align}
\label{main control}
|0\rangle_0
|\psi(\phi_1)\rangle_1
|0\rangle_2
|\Psi_c(\phi_c)\rangle_3
&\xrightarrow{((g \leftrightarrow ME3) +\pi)}
\nonumber  \\
|\tilde\Psi_c(\phi_c+\pi )\rangle_0
&|\psi(\phi_1)\rangle_1
|0\rangle_2
|0\rangle_3.
\end{align}

\noindent
where $|\tilde\Psi_c(\phi_c+\pi )\rangle_0 =\{\alpha_c|0\rangle_0-\beta_c e^{i\phi_c}|b\rangle_0\}$
Now we realize the scenario of resonant (or nonresonant) swapping for the state (\ref{main control})

\begin{align}
\label{plus SWAP}
&|\tilde\Psi_c(\phi_c+\pi )\rangle_0
|\psi(\phi_1)\rangle_1
|0\rangle_2 |0\rangle_3
\xrightarrow{((g \leftrightarrow ME3) +\pi) + swap}
\nonumber  \\
&\{\alpha_c |0\rangle_0  |0\rangle_1 |\psi(\phi_1-\tilde \omega t^{(r,nr)}_{swap}+\pi )\rangle_2
\nonumber  \\
&-\beta_c e^{i\phi_c}|b\rangle_0
|\psi(\phi_1)\rangle_1
|0\rangle_2 \}|0\rangle_3.
\end{align}

\noindent
After subsequent using of the laser $\pi$-pulse on the transition $|1\rangle_0\leftrightarrow|b\rangle_0$ and resonant transfer between ME3-node and \emph{gate-}atoms, we terminate a realization of $SWAP$-gate controlled by the qubit state of ME3-node

\begin{align}
\label{control SWAP}
& |0\rangle_0
|\psi(\phi_1)\rangle_1
|0\rangle_2
|\Psi_c(\phi_c+\pi )\rangle_3
\xrightarrow{control-SWAP}
\nonumber  \\
&|0\rangle_0  \{\alpha_c |0\rangle_1
|\psi(\phi_1-\tilde \omega t^{(r,nr)}_{swap}+\pi )\rangle_2 |0\rangle_3
\nonumber  \\
&+\beta_c e^{i\phi_c}
|\psi(\phi_1)\rangle_1
|0\rangle_2 |1\rangle_3\}.
\end{align}

\noindent
As seen here, AQT operation leads to entanglement  of ME3-node with ME1 and ME2-nodes coupled via $SWAP$-gate
where the entanglement gets a maximum at $|\alpha_c|=|\beta_c|=1/\sqrt{2}$.
Similarly we realize a $\sqrt{SWAP}$-gate controlled by ME3-node

\begin{align}
\label{control sqrt SWAP}
&|0\rangle_0
|\psi(\phi_1)\rangle_1
|0\rangle_2
|\Psi_c(\phi_c+\pi )\rangle_3
\xrightarrow{control-\sqrt{swap}}
\nonumber  \\
&|0\rangle_0 \{\alpha_c   |\Phi^{(nr)}_{\sqrt{swap}}(\psi)\rangle |0\rangle_3
+\beta_c e^{i\phi_c}
|\psi(\phi_1)\rangle_1
|0\rangle_2 |1\rangle_3\},
\nonumber  \\
\end{align}

\noindent
 and another control-$SWAP(\theta_{n,m} )$ gate that yields more sophisticated quantum superposition coupling the entangled state caused by $SWAP(\theta_{n,m} )$ gate (for example, state (\ref{function of sqrt_SWAP}) for $\sqrt{SWAP}$)  and initial qubit state of ME1-node.

\emph{Discussion and Conclusion: }
We have described main properties of the proposed atomic quantum transistor (AQT). Here, we have found novel effect of quantum transport between resonant atomic nodes providing a perfect realization of control-$SWAP(\theta_{n,m})$ processes.
The elaborated model of the quantum transport can be implemented with various quantum systems (natural and artificial atoms and molecules) by using  nano-optical \cite{Vetsch2010} and nano-plasmonic \cite{Chang2007,Kamli2008} schemes interesting for quantum computing and communication.
It is worth noting that by using the  swapping processes, one can realize a complete set of single and two-qubit gates in a system of many atomic nodes where $SWAP$- and control-$SWAP$ processes will play a role of single- and two-qubit gates.
Detailed study of these quantum protocols will be a subject of further investigation.
From all set of possible quantum states, here we briefly discuss only one specific eigenstate (we call it \emph{spatial dark multi-atomic} (\emph{SDMA-}) state $|\Psi_{sdma}\rangle$) of \emph{gate-}atom and ME1, ME2-nodes coupled by the effective Hamiltonian (\ref{main Hamiltonian1}).  SDMA-state can be  prepared via $\sqrt{SWAP}-$ process in Eq. (\ref{function of sqrt_SWAP}) with additional $\pi/2-$phase shift of atoms in ME2-node caused by using a local Stark or Zeeman shift of the atomic levels during short enough time. This procedure leads to the state $|\Psi_{sdma} \rangle =(|\Psi_{1} \rangle -|\Psi_{2} \rangle)/{\sqrt{2}}=|0\rangle_0 (|10 \rangle - |01\rangle)/{\sqrt{2}}$ which demonstrates an immunity to the
exchange by virtual photons between three atomic groups  as in the case of usual atomic dark state \cite{Fleischhauer2005} that demonstrates a complete decoupling from the interaction with two (or more) coherent light field in the electromagnetically induced  transparency (EIT) effect. By taking into account that SDMA-state belongs to the decoherence free subspace for quantum dynamics of the atoms, this points out an important resource for realization of long-lived quantum coherence in this kind of multi-atomic systems evolving in nano-optical scheme.
SDMA-state and state (\ref{main solution-c}) (which is a \emph{spatial bright multi-atomic} state in contrast to SDMA-state) are a clear demonstration of the entanglement in many atomic systems situated in the distant nodes. The states reveal themselves in a new effect of spatial quantum interference at coherent transport through our three-node chain similarly to dark and bright states in quantum transport based on EIT effect.

Financial support by the Russian Fund for Basic Research grant \#
10-02-01348-a is gratefully acknowledged.

\end{document}